\documentstyle[preprint,aps,floats,tighten,epsf]{revtex}
\begin{document}

\preprint{Fermilab-Pub-97/273-E}

\title{Experimental Search for Chargino and Neutralino Production
       in Supersymmetry Models with a Light Gravitino}

\author{
\centerline{The D\O\ Collaboration\thanks{Authors listed on the following page.
            \hfill\break
            To be Published in Physical Review Letters.}}
}
\address{
\centerline{Fermi National Accelerator Laboratory, Batavia, Illinois 60510}
}

\date{August 4, 1997}

\maketitle

\begin{abstract}
  We search for inclusive high $E_T$ diphoton events
  with large missing transverse energy
  in $p\bar{p}$\ collisions at $\sqrt{s}$=1.8~TeV.  Such events are expected
  from pair production of charginos and neutralinos within the framework
  of the minimal supersymmetric standard model with a light gravitino. No
  excess of events is observed. In that model, and assuming gaugino mass
  unification at the GUT scale, we obtain a 95\% CL exclusion region in the
  supersymmetry parameter space and lower mass bounds
  of 150~GeV/c$^2$ for the lightest chargino and 75~GeV/c$^2$ for the
  lightest neutralino.
\end{abstract}


\newpage
\begin{center}
B.~Abbott,$^{28}$                                                             
M.~Abolins,$^{25}$                                                            
B.S.~Acharya,$^{43}$                                                          
I.~Adam,$^{12}$                                                               
D.L.~Adams,$^{37}$                                                            
M.~Adams,$^{17}$                                                              
S.~Ahn,$^{14}$                                                                
H.~Aihara,$^{22}$                                                             
G.A.~Alves,$^{10}$                                                            
E.~Amidi,$^{29}$
% ...
\end{center}
\vskip 0.50cm
\normalsize                                                                 
\centerline{(D\O\ Collaboration)}                                             

\newpage
\small
\it
\centerline{$^{1}$Universidad de los Andes, Bogot\'{a}, Colombia}             
\centerline{$^{2}$University of Arizona, Tucson, Arizona 85721}               
\centerline{$^{3}$Boston University, Boston, Massachusetts 02215}             
\centerline{$^{4}$Brookhaven National Laboratory, Upton, New York 11973}      
\centerline{$^{5}$Brown University, Providence, Rhode Island 02912}           
\centerline{$^{6}$Universidad de Buenos Aires, Buenos Aires, Argentina}       
\centerline{$^{7}$University of California, Davis, California 95616}          
\centerline{$^{8}$University of California, Irvine, California 92697}         
\centerline{$^{9}$University of California, Riverside, California 92521}      
\centerline{$^{10}$LAFEX, Centro Brasileiro de Pesquisas F{\'\i}sicas,        
                  Rio de Janeiro, Brazil}                                     
\centerline{$^{11}$CINVESTAV, Mexico City, Mexico}                            
\centerline{$^{12}$Columbia University, New York, New York 10027}             
\centerline{$^{13}$Delhi University, Delhi, India 110007}                     
\centerline{$^{14}$Fermi National Accelerator Laboratory, Batavia,            
                   Illinois 60510}                                            
\centerline{$^{15}$Florida State University, Tallahassee, Florida 32306}      
\centerline{$^{16}$University of Hawaii, Honolulu, Hawaii 96822}              
\centerline{$^{17}$University of Illinois at Chicago, Chicago,                
                   Illinois 60607}                                            
\centerline{$^{18}$Indiana University, Bloomington, Indiana 47405}            
\centerline{$^{19}$Iowa State University, Ames, Iowa 50011}                   
\centerline{$^{20}$Korea University, Seoul, Korea}                            
\centerline{$^{21}$Kyungsung University, Pusan, Korea}                        
\centerline{$^{22}$Lawrence Berkeley National Laboratory and University of    
                   California, Berkeley, California 94720}                    
\centerline{$^{23}$University of Maryland, College Park, Maryland 20742}      
\centerline{$^{24}$University of Michigan, Ann Arbor, Michigan 48109}         
\centerline{$^{25}$Michigan State University, East Lansing, Michigan 48824}   
\centerline{$^{26}$Moscow State University, Moscow, Russia}                   
\centerline{$^{27}$University of Nebraska, Lincoln, Nebraska 68588}           
\centerline{$^{28}$New York University, New York, New York 10003}             
\centerline{$^{29}$Northeastern University, Boston, Massachusetts 02115}      
\centerline{$^{30}$Northern Illinois University, DeKalb, Illinois 60115}      
\centerline{$^{31}$Northwestern University, Evanston, Illinois 60208}         
\centerline{$^{32}$University of Notre Dame, Notre Dame, Indiana 46556}       
\centerline{$^{33}$University of Oklahoma, Norman, Oklahoma 73019}            
\centerline{$^{34}$University of Panjab, Chandigarh 16-00-14, India}          
\centerline{$^{35}$Institute for High Energy Physics, 142-284 Protvino,       
                   Russia}                                                    
\centerline{$^{36}$Purdue University, West Lafayette, Indiana 47907}          
\centerline{$^{37}$Rice University, Houston, Texas 77005}                     
\centerline{$^{38}$Universidade do Estado do Rio de Janeiro, Brazil}          
\centerline{$^{39}$University of Rochester, Rochester, New York 14627}        
\centerline{$^{40}$CEA, DAPNIA/Service de Physique des Particules,            
                   CE-SACLAY, Gif-sur-Yvette, France}                         
\centerline{$^{41}$Seoul National University, Seoul, Korea}                   
\centerline{$^{42}$State University of New York, Stony Brook,                 
                   New York 11794}                                            
\centerline{$^{43}$Tata Institute of Fundamental Research,                    
                   Colaba, Mumbai 400005, India}                              
\centerline{$^{44}$University of Texas, Arlington, Texas 76019}               
\centerline{$^{45}$Texas A\&M University, College Station, Texas 77843}       
\normalsize

\newpage


Supersymmetric models with a light gravitino~($\tilde G$), 
first proposed by Fayet\cite{fayet}, have generated recent
theoretical interest\cite{all,eln,tanb}.
These models are characterized by a supersymmetry breaking scale
$\Lambda$ as low as 100~TeV and a gravitino which
is naturally the lightest supersymmetric particle~(LSP). 
The lightest superpartner of a standard model particle,
assumed here and in most analyses to be the
lightest neutralino~($\tilde\chi^0_1$),
is the next-to-lightest supersymmetric particle (NLSP).
If $\tilde\chi^0_1$ has a non-zero photino component,
it is unstable and decays into a photon
plus a gravitino ($\tilde\chi^0_1\rightarrow\gamma\tilde G$).

In this Letter, we present a direct search for supersymmetry with a light
gravitino
in the framework of the minimal supersymmetric standard model~(MSSM).
In this framework the
gaugino-Higgsino sector (excluding gluinos) is described by
four parameters: $M_1$, $M_2$,
$\mu$, and $\tan\beta$, where $M_1$ and $M_2$ are the $U(1)$ and $SU(2)$
gaugino mass parameters, $\mu$ is the Higgsino mass
parameter, and $\tan\beta$ is the ratio of the vacuum expectation values
of the two Higgs doublets\cite{susy}.  
With the assumption of gaugino mass unification
at the GUT scale,
$M_1 = \frac{5}{3} M_2\tan^2\theta_{\it W}$,
where $\theta_{\it W}$ is the weak mixing angle. There are four neutralinos
($\tilde\chi^0_i,\ i=1,2,3,4$) and two charginos~($\tilde\chi^\pm_j,\ j=1,2$)
whose masses and couplings are fixed by $M_2$, $\mu$ and $\tan\beta$.
We assume $\tan\beta>1$ in this analysis.

We search for neutralino and chargino
pair production in $\sqrt{s}$=1.8~TeV
$p\bar{p}$\ collisions at the Fermilab Tevatron.
The $\tilde\chi^0_1$ is assumed to be short-lived,
decaying within the detector to $\gamma\tilde G$ with a branching ratio of
100\%.  Decay to a Higgs boson is assumed to be kinematically inaccessible.
R-parity conservation is assumed so that supersymmetric particles
are pair produced and the LSP is stable and
non--interacting.  Thus pair production of
charginos and neutralinos yields
$\gamma\gamma\rlap{\kern0.25em/}E_T$\ events with high transverse
energy ($E_T$) photons and
large missing transverse energy~($\rlap{\kern0.25em/}E_T$),
with or without jets.
The high $E_T$ photons and large $\rlap{\kern0.25em/}E_T$\
provide a powerful tool
for suppressing backgrounds.

Recently D\O\ reported a search\cite{johnw} for
$\gamma\gamma\rlap{\kern0.25em/}E_T$\ events
based on supersymmetry models
with $\tilde\chi^0_1$ as the LSP.  In this analysis, we present the first
experimental study of
$p\bar{p}\rightarrow\gamma\gamma\rlap{\kern0.25em/}E_T + X$
based on the MSSM with a light gravitino
as the LSP.  Using this model and more efficient photon
identification and event selection
criteria than in Ref.\cite{johnw}, we set the strongest limits to
date in the supersymmetry parameter space, exceeding those from 
LEP experiments\cite{eln}.


The data used in this analysis were collected with the D\O\ detector during the
1992--1996 Tevatron run at $\sqrt{s}$=1.8~TeV and represent
an integrated luminosity of $106.3\pm 5.6$~pb$^{-1}$.
A detailed description of the D\O\
detector can be found in Ref.\cite{dzero}. The trigger requires
one electromagnetic (EM) cluster with transverse energy $E_T>15$~GeV,
one jet with $E_T>10$~GeV, and $\rlap{\kern0.25em/}E_T$$>14$~GeV
($\rlap{\kern0.25em/}E_T$$>10$~GeV for about 10\% of
the data taken early in the Tevatron run). 
The jets in the trigger include non-leading EM clusters.
Photons are identified through
a two-step process: the selection of isolated EM energy
clusters and
the rejection of electrons. The EM clusters are selected from
calorimeter energy clusters by requiring
(i)   at least 95\% of the energy to be deposited in the EM section
      of the calorimeter,
(ii)  the transverse and longitudinal shower profiles to be consistent
      with those expected for an EM shower, and
(iii) the energy in an annular isolation cone from radius 0.2 to 0.4 around
      the cluster in $\eta-\phi$ space to be less than 10\% of the cluster
      energy, where $\eta$ and $\phi$ are the pseudorapidity and azimuthal
      angle.
Electrons are removed by rejecting EM clusters
which have either a reconstructed track or a large number of tracking chamber
hits in a road between the calorimeter cluster and the event
vertex. $\rlap{\kern0.25em/}E_T$~is determined
from the energy deposition in the calorimeter for $|\eta|<4.5$.

To be selected as $\gamma\gamma\rlap{\kern0.25em/}E_T$\ candidates, events
are first required to have two identified photons,
one with $E_T^{\gamma_1}>20$~GeV and the other with $E_T^{\gamma_2}>12$~GeV,
each with pseudorapidity $|\eta^{\gamma}|<1.2$
or $1.5<|\eta^{\gamma}|<2.0$.  
We denote the 28 events passing these photon requirements
as the $\gamma\gamma$ sample.
We then require $\rlap{\kern0.25em/}E_T$$>25$~GeV
with at least one reconstructed vertex in the event
to ensure good measurement of $\rlap{\kern0.25em/}E_T$. No requirement
on jets is made.
Two events satisfy all requirements.

%
%

The principal backgrounds are
multijet, direct photon, $W+\gamma$, $W+{\rm jets}$,
$Z\rightarrow ee$, and $Z\rightarrow\tau\tau\rightarrow ee$ events from
Standard Model processes with misidentified photons and/or mismeasured
$\rlap{\kern0.25em/}E_T$.  The background due
to $\rlap{\kern0.25em/}E_T$\ mismeasurement
is estimated using events with two EM-like clusters 
which satisfy looser EM cluster requirements than those discussed
above, and for which at least one of the two fails the 
EM shower profile consistency requirement (ii) above. 
In addition, these events must pass the photon kinematic
requirements.  These events, called the QCD sample, are similar
to those of the $\gamma\gamma$ sample and are expected to have
similar $\rlap{\kern0.25em/}E_T$\ resolution.
By normalizing the number of events with
$\rlap{\kern0.25em/}E_T$\ $<20$~GeV in the
QCD sample to that in the $\gamma\gamma$ sample,
we obtain a background of $2.1\pm 0.9$ events due
to $\rlap{\kern0.25em/}E_T$\ mismeasurement
for $\rlap{\kern0.25em/}E_T$\ $>25$~GeV.

\begin{figure}
  \centerline{\epsfysize=3.2in \epsfbox{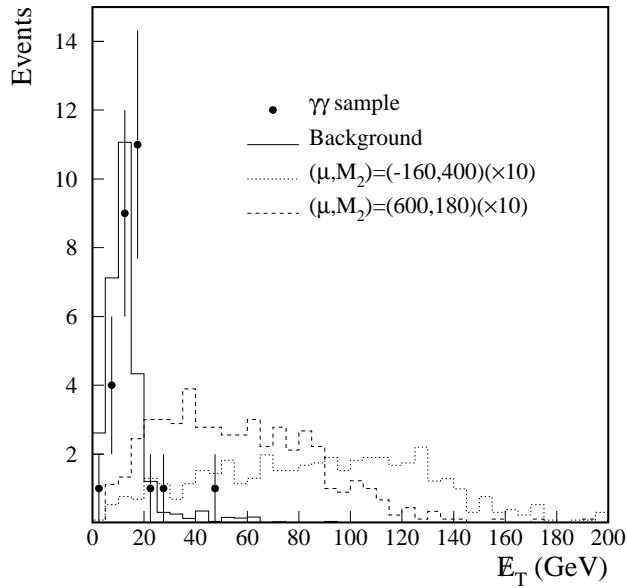}}
    \caption{The \protect$\rlap{\kern0.25em/}E_T$\
             distributions of the $\gamma\gamma$ and
             background samples. The number of events with
             \protect$\rlap{\kern0.25em/}E_T$$<20$~GeV in the background
             sample is normalized
             to that in the $\gamma\gamma$ sample. Also shown are the
             expected distributions (multiplied by 10)
             from two representative points in
             the supersymmetry parameter space, with $\tan\beta=2$.}
    \label{fig:prl1}
\end{figure}

Other backgrounds are due to events with genuine $\rlap{\kern0.25em/}E_T$\
such as those from
$W+$`$\gamma$' (where `$\gamma$' can be a real or a fake photon),
$Z\rightarrow\tau\tau\rightarrow ee$, and
$t\bar{t}\rightarrow ee+{\rm jets}$ production. These events
would fake $\gamma\gamma\rlap{\kern0.25em/}E_T$\ events if the electrons
were misidentified as photons. We estimate their contribution using a
sample of $e$`$\gamma$' events
passing the kinematic requirements, including that on $\rlap{\kern0.25em/}E_T$.
Electrons are selected from the identified EM clusters
with matched tracks.  
Taking into account the probability (0.0045) that an electron is misidentified
as a photon, we 
estimate a background of $0.2\pm 0.1$ events.
Adding the two background contributions together yields $2.3\pm 0.9$ events.
The $\rlap{\kern0.25em/}E_T$\ distributions of
the $\gamma\gamma$ sample and the background sample
are compared in Fig.~\ref{fig:prl1}.  Also shown are
the expected distributions
from supersymmetry for two representative points in the $(\mu,M_2)$
parameter space.

%
%
Chargino and neutralino pair production and
decay are modeled using the
{\sc spythia}\ program\cite{spythia}, a supersymmetric extension of the
{\sc pythia~5.7}\ program\cite{pythia}.  Squarks and sleptons
are assumed to be heavy.  This assumption is conservative because light 
sleptons would lead to events with less jet activity and would therefore 
improve detection efficiency.  For light squarks, no change in efficiency is
expected.
To explore the parameter space, we choose to work in the $(\mu,M_2)$ plane 
while keeping $\tan\beta$ fixed. We generate
$\tilde\chi^0_i\tilde\chi^0_j$,\ $\tilde\chi^0_i\tilde\chi^\pm_j$\ and
$\tilde\chi^\pm_i\tilde\chi^\pm_j$\
events for a large number of points in the $(\mu, M_2)$ parameter space.
Table~\ref{tab:tab1} shows the resulting theoretical cross sections
$\sigma_{th}$ for several representative points,
calculated using the
CTEQ3L parton distribution function\cite{cteq}.
To determine the signal efficiencies,
Monte Carlo events are run through a {\sc geant}\ \cite{geant}
based D\O\ detector simulation program, a trigger simulator, and the same
trigger requirements, reconstruction, and analysis as the data.
The total signal efficiency $\epsilon$ (including efficiencies of the
trigger, reconstruction, photon identification, and kinematic requirements)
varies greatly, from $\sim 0.01\%$ to $\sim 26\%$,
depending largely on the masses of $\tilde\chi^\pm_1$ and $\tilde\chi^0_1$
and their mass difference.  For large masses such as
those in Table~\ref{tab:tab1}, the total efficiency $\epsilon > 15\%$.
The estimated systematic error on the total
efficiency is $0.06\epsilon$.


With two events observed and $2.3\pm 0.9$ events expected from background,
we observe no excess of events.
We compute 95\% CL upper limits on the cross section $\sigma$
for the Monte Carlo sampled points in the ($\mu, M_2$)
plane using a Bayesian approach\cite{johnh} with a flat prior distribution
for the signal cross section. The calculation takes into account the errors
on the luminosity, the efficiency, and the number of background events.
Depending on the values of the supersymmetry parameters, the 95\% CL
upper limits on the total cross section vary
widely from several hundred pb for light charginos/neutralinos to
$\sigma \sim 0.18$ pb for heavy charginos/neutralinos.
The upper limit $\sigma_D$ quoted in Table~\ref{tab:tab1}
is for events satisfying the kinematic cuts of this analysis
at the generator level;
comparison of $\sigma$ and $\sigma_D$ indicates the fraction of events 
yielding detectable particles for the various parameter points.

\begin{table}
\squeezetable
    \begin{tabular}{cc|cc|c|cc|cc}
 $\mu$  & $M_2$ & $m_{\tilde\chi^0_1}$ & $m_{\tilde\chi^\pm_1}$ &
 $\sigma_{th}$ &  \multicolumn{2}{c|}{Efficiencies (\%)} &
 \multicolumn{2}{c}{Limits (pb)} \\
  GeV  & GeV &\multicolumn{2}{c|}{GeV/c$^2$} & (pb) &
  $\epsilon$ & $\epsilon_D$ &
  $\sigma$ & $\sigma_D$ \\ \hline
   -160 & 300 & 143.9 & 167.8 & 0.12 & 26.0$\pm$1.4 & 36.4 & 0.18 & 0.13 \\
   -600 & 140 &  72.5 & 146.4 & 0.36 & 17.2$\pm$1.2 & 32.1 & 0.28 & 0.15 \\
   -800 & 165 &  84.7 & 170.0 & 0.20 & 15.1$\pm$1.1 & 26.4 & 0.32 & 0.18 \\
    200 & 300 & 118.1 & 160.2 & 0.15 & 21.3$\pm$1.3 & 31.9 & 0.23 & 0.15 \\
    400 & 190 &  89.4 & 166.4 & 0.19 & 20.1$\pm$1.3 & 32.5 & 0.24 & 0.15 \\
    800 & 170 &  83.2 & 161.6 & 0.25 & 19.6$\pm$1.3 & 33.4 & 0.25 & 0.14 \\
    \end{tabular}
    \caption{Representative points in the ($\mu,M_2$) plane for $\tan\beta=2$
             with GEANT simulation.  These points are chosen to be near our
             95\% CL bounds, where the experimental 95\% CL
             cross section $\sigma$ equals the theoretical cross section
             $\sigma_{th}$. The efficiency $\epsilon$ is for
             observing the total cross section $\sigma$, while
             $\epsilon_D$ and $\sigma_D$ are the efficiency and cross section
             for observing the detectable events, those
             which satisfy the kinematic cuts $E_T^{\gamma_1}>20$~GeV,
             $E_T^{\gamma_2}>12$~GeV, $|\eta^{\gamma}|<1.2$
             or $1.5<|\eta^{\gamma}|<2.0$, and
             $\rlap{\kern0.25em/}E_T >25$~GeV at the generator level.  
             The total efficiency $\epsilon = \epsilon_D \times \epsilon_K$ 
             where $\epsilon_K$ is the
             efficiency of the kinematic cuts.}
  \label{tab:tab1}
\end{table}

\begin{figure}
    \centerline{\epsfysize=3.2in \epsfbox{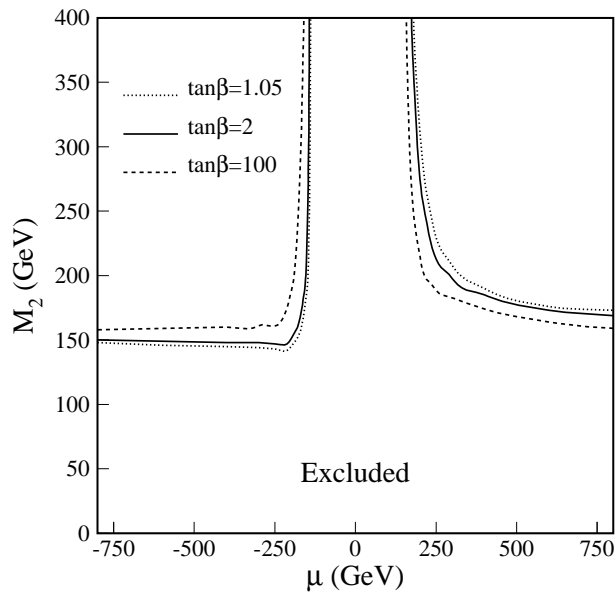}}
    \caption{95\% CL bounds in the ($\mu,M_2$) plane for $\tan\beta=2$
             (solid line), $\tan\beta=1.05$ (dotted line), and $\tan\beta=100$
             (dashed line).}
    \label{fig:prl2}
\end{figure}

To derive bounds in the ($\mu,M_2$) plane, the values of $\mu$ and $M_2$ are
varied around the sampled points until the
theoretical cross sections $\sigma_{th}$ exceed
the upper limits $\sigma$. The interpolated bounds in the
$(\mu,M_2)$ plane are shown in Fig.~\ref{fig:prl2} for $\tan\beta=1.05,2,100$.
The regions below the lines are excluded by this analysis.
The bounds depend on the value of $\tan\beta$ slightly,
becoming stronger in the $\mu<0$ half-plane and weaker in the other
half-plane as $\tan\beta$ is increased.

\begin{figure}
    \centerline{\epsfysize=3.2in \epsfbox{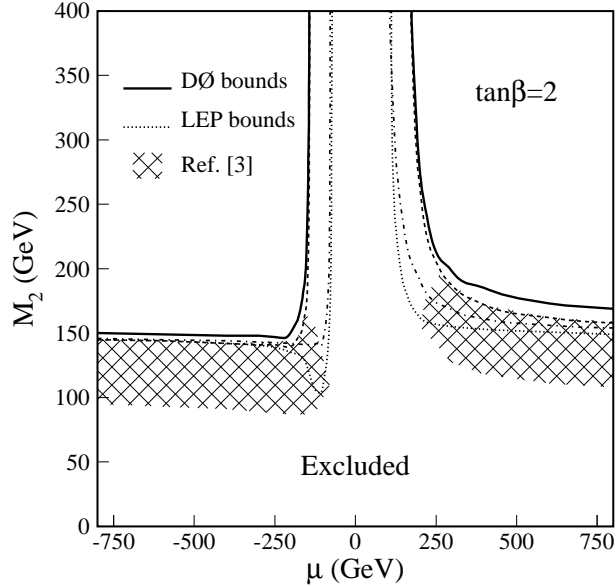}}
    \caption{Bounds in the ($\mu,M_2$) plane for
             $\tan\beta=2$. The region below the two solid lines is excluded
             at 95\% CL. Also shown are the bounds estimated in
             Ref.~\protect\cite{eln} from LEP data 
             (dotted line) and the contours
             of constant $m_{\tilde\chi^\pm_1}=150$~GeV/c$^2$ (dashed line)
             and $m_{\tilde\chi^0_1}=75$~GeV/c$^2$ (dot-dashed line).
             The hatched areas are suggested in Ref.~\protect\cite{eln}
             for the chargino interpretation of the CDF event candidate
             in the model.}
    \label{fig:prl3}
\end{figure}

Figure~\ref{fig:prl3} compares
the bounds in the ($\mu,M_2$) plane for $\tan\beta=2$ with those
estimated from LEP data\cite{eln} within the
framework of a light gravitino
and assuming a 75 GeV/c$^2$ selectron for
t-channel exchange.  These bounds
exclude the region of parameter space suggested in Ref.\cite{eln} for
the chargino interpretation of an event candidate shown by the
CDF Collaboration\cite{cdf}.
Also shown are
the contours of constant mass for $m_{\tilde\chi^\pm_1}=150$~GeV/c$^2$ and
$m_{\tilde\chi^0_1}=75$~GeV/c$^2$. Since these are the largest masses for which
the mass contours lie entirely in the excluded region, we obtain
95\% CL lower mass limits of 150~GeV/c$^2$ for the
lightest chargino and 75~GeV/c$^2$ for the lightest neutralino.
This 75~GeV/c$^2$ lower mass limit also rules out a large part of the parameter
space suggested for the selectron interpretation
of the CDF event candidate in the model, as discussed in Ref.\cite{eln}.
These mass limits are insensitive to
the choice of $\tan\beta$, varying less
than 2~GeV/c$^2$ over the range $1.05<\tan\beta< 100$,
as long as our assumption that
$\tilde\chi^0_1$ is the NLSP is satisfied.
For large $\tan\beta$ values, 
this assumption may not be satisfied~\cite{tanb}.

\begin{figure}
    \centerline{\epsfysize=3.2in \epsfbox{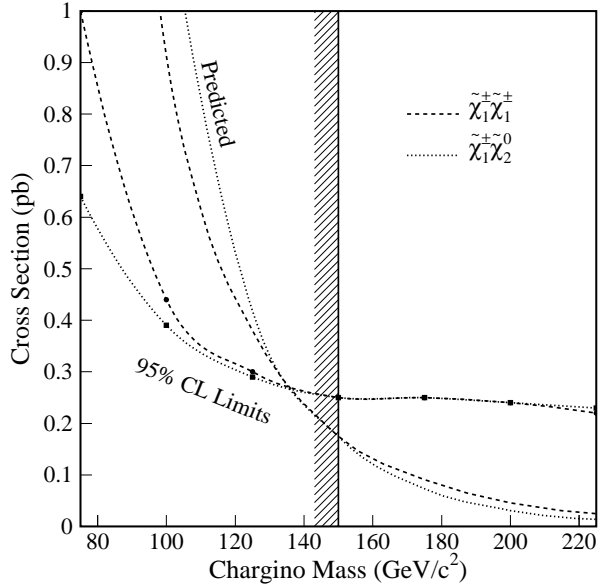}}
    \caption{Measured 95\% CL upper limits and predicted theoretical cross
             sections for $\tilde\chi^\pm_1\tilde\chi^\pm_1$
             and  $\tilde\chi^\pm_1\tilde\chi^0_2$ production as a function of
             $m_{\tilde\chi^\pm_1}$, assuming $m_{\tilde\chi^\pm_1}\approx
             m_{\tilde\chi^0_2}\approx 2\times m_{\tilde\chi^0_1}$.
             The vertical hatched line is the 95\% CL lower limit on
             $m_{\tilde\chi^\pm_1}$ determined using the total
             cross section for all chargino/neutralino pair production
             and all possible $\mu$ values, as determined in this paper.}
    \label{fig:prl4}
\end{figure}

Most of the theoretical cross section for the 
$\gamma\gamma\rlap{\kern0.25em/}E_T$\ process
is due to
$\tilde\chi^\pm_1\tilde\chi^\pm_1$ and
$\tilde\chi^\pm_1\tilde\chi^0_2$ production.  
For the large part of the parameter space with $|\mu | > M_2$,
the relation 
$m_{\tilde\chi^\pm_1}\approx m_{\tilde\chi^0_2}\approx 2\times
m_{\tilde\chi^0_1}$ holds, so we can express our
cross section limits simply in terms of $m_{\tilde\chi^\pm_1}$.
Figure~\ref{fig:prl4} shows the 95\% CL upper limits for 
both processes, together with the theoretical predictions for
$\tan\beta=2$ and $\mu=-500$~GeV.  The experimental limits are
insensitive to the choice of $\tan\beta$ and $\mu$ while the theoretical
cross section varies by about 10\%.  
Our data rule out chargino masses below $\approx 137$~GeV/c$^2$ in 
models with a light gravitino,
assuming $|\mu | > M_2$.
This limit, though weaker than the 
150~GeV/c$^2$ limit determined above from all processes 
contributing to 
$\gamma\gamma\rlap{\kern0.25em/}E_T$\ final states, is useful for
comparison with semi-exclusive calculations of gaugino production.

In summary,
we have searched for inclusive high $E_T$ diphoton events with large missing
transverse energy.  
Such events are expected in the framework of supersymmetric
models with a light gravitino.
No excess of
events is found.  The null result, interpreted in this framework 
and with the assumption
of gaugino mass unification at the GUT scale, yields 
a 95\% CL lower mass limits of 150~GeV/c$^2$ for the
lightest chargino and 75~GeV/c$^2$ for the lightest neutralino.


We thank G.L.~Kane for many useful discussions
and S.~Mrenna for his help in using the {\sc spythia}\ program.
We also thank the staffs at Fermilab and collaborating institutions for their
contributions to this work, and acknowledge support from the
Department of Energy and National Science Foundation (U.S.A.),
Commissariat  \` a L'Energie Atomique (France),
State Committee for Science and Technology and Ministry for Atomic
   Energy (Russia),
CNPq (Brazil),
Departments of Atomic Energy and Science and Education (India),
Colciencias (Colombia),
CONACyT (Mexico),
Ministry of Education and KOSEF (Korea),
and CONICET and UBACyT (Argentina).



\begin{thebibliography}{99}
%
\bibitem[*]{beijing}
Visitor from IHEP, Beijing, China.

\bibitem[\dag]{ecuador}
Visitor from Universidad San Francisco de Quito, Quito, Ecuador.

\vskip 0.25cm

  \bibitem{fayet}
     P. Fayet, Phys. Lett. B {\bf 70}, 461 (1977);
               {\it ibid.} B {\bf 86}, 272 (1979);
               {\it ibid.} B {\bf 175}, 471 (1986);
     M. Dine, W. Fischler and M. Srednicki, Nucl. Phys. B {\bf 189}, 575 (1981);
     S. Dimopoulos and S. Raby, Nucl. Phys. B {\bf 192}, 353 (1981);
     D.A. Dicus, S. Nandi, and J. Woodside, Phys. Rev. D {\bf 41}, 2347 (1990);
                                         {\it ibid.} D {\bf 43}, 2951 (1991);
     M. Dine {\it et al.}, Phys. Rev. D {\bf 53}, 2658 (1996).

  \bibitem{all}
     D.R. Stump, M. Wiest, and C.P. Yuan, Phys. Rev. D {\bf 54}, 1936 (1996);
     S. Dimopoulos, S. Thomas, and J.D. Wells,
                   Phys. Rev. D {\bf 54}, 3283 (1996);
     S. Dimopoulos {\it et al.}, Phys. Rev. Lett. {\bf 76}, 3494 (1996);
     S. Ambrosanio {\it et al.}, Phys. Rev. Lett. {\bf 76}, 3498 (1996),
                   Phys. Rev. D {\bf 54}, 5395 (1996);
     K.S. Babu, C. Kolda, and F. Wilczek,
                   Phys. Rev. Lett. {\bf 77}, 3070 (1996);
     J.L. Lopez, D.V. Nanopoulos, and A. Zichichi,
                  Phys. Rev. Lett. {\bf 77}, 5168 (1996) and
                                               hep-ph/9610235 (unpublished).

  \bibitem{eln}
     J. Ellis, J.L. Lopez, and D.V. Nanopoulos,
                             Phys. Lett. B {\bf 394}, 354 (1997).

  \bibitem{tanb}
     H. Baer {\it et al.}, Phys. Rev. D {\bf 55}, 4463 (1997);
     S. Ambrosanio, G. Kribs and S. Martin, Phys. Rev. D {\bf 56}, 1761 (1997).

  \bibitem{susy}
     H. E. Haber and G. Kane, Phys. Rept. {\bf 117}, 75 (1985).

  \bibitem{johnw}
     D\O\ Collaboration, S. Abachi {\it et al.},
                                Phys. Rev. Lett. {\bf78}, 2070 (1997).

  \bibitem{dzero}
     D\O\ Collaboration, S. Abachi {\it et al.}, Nucl. Instrum. Methods
                         A {\bf 338}, 185 (1994).

  \bibitem{spythia}
     S. Mrenna, hep-ph/9609360 (unpublished).

  \bibitem{pythia}
     H.-U. Bengtsson and T. Sj\"ostrand, Comp. Phys. Comm. {\bf 46}, 43 (1987);
     T. Sj\"ostrand, Comp. Phys. Comm. {\bf 82}, 74 (1994).

  \bibitem{cteq}
     H. L. Lai {\it et al.}, Phys. Rev. D {\bf 51}, 4763 (1995).

  \bibitem{geant}
     R. Brun and F. Carminati, CERN Program Library Long Writeup W5013,
                               1993 (Unpublished).

  \bibitem{johnh}
     Particle Data Group, R. M. Barnett {\it et al.},
               Phys. Rev. D {\bf 54}, 1 (1996).

  \bibitem{cdf}
     S. Park, ``{\em Search for New Phenomena in CDF}'',
                Proceedings of the
                10th Topical Workshop on Proton-Antiproton Collider Physics,
                Batavia, 1995, edited by R. Raja and J. Yoh 
                (AIP Press, 1995), p. 62.

\end{thebibliography}
\end{document}